# Photostrictive actuators based on freestanding ferroelectric membranes

*Saptam Ganguly[1] \*, David Pesquera[1] \*, Daniel Moreno Garcia[3], Umair Saeed[1], Nona Mirzamohammadi[1], José Santiso[1], Jessica Padilla[1], José Manuel Caicedo Roque[1], Claire Laulhé[4], Felisa Berenguer[5], Luis Guillermo Villanueva[3], Gustau Catalan[1,2] \**

[1]Catalan Institute of Nanoscience and Nanotechnology (ICN2), CSIC and BIST, Campus UAB, Bellaterra, Barcelona, Catalonia, Spain.

[2]ICREA - Institució Catalana de Recerca i Estudis Avançats, Barcelona, Catalonia.

[3]Advanced NEMS Laboratory, Institute of Mechanical Engineering, École Polytechnique Fédérale de Lausanne (EPFL), 1015, Lausanne, Switzerland.

[4] Université Paris-Saclay, Synchrotron SOLEIL, 91190 Saint-Aubin, France.

[5] Synchrotron SOLEIL, L'Orme des Merisiers, Saint-Aubin BP 48, 91190 Gif-sur-Yvette, France.

**Abstract** Complex oxides offer a wide range of functional properties, and recent advances in fabrication of freestanding membranes of these oxides are adding new mechanical degrees of freedom to this already rich functional ecosystem. Here, we demonstrate photoactuation in freestanding thin film resonators of ferroelectric Barium Titanate ($BaTiO_3$) and paraelectric Strontium Titanate ($SrTiO_3$). The free-standing films, transferred onto perforated supports, act as nano-drums, oscillating at their natural resonance frequency when illuminated by a frequency-modulated laser. The light-induced deflections in the ferroelectric $BaTiO_3$ membranes are two orders of magnitude larger than in the paraelectric $SrTiO_3$ ones. Time-resolved X-ray micro-diffraction under illumination and temperature-dependent and holographic interferometry provide combined evidence for the photostrictive strain in $BaTiO_3$ originating from partial screening of ferroelectric polarization by photo-excited carriers, which decreases the tetragonality of the unit cell. These findings showcase the potential of photostrictive freestanding ferroelectric films as wireless actuators operated by light.

# 1    Introduction

Ferroelectric materials deform under voltage, owing to the piezoelectric coupling between polarization and strain. This electromechanical coupling has been exploited in applications such as sensors, actuators, and energy harvesters,[1–5] and integrated in micro/nano-electromechanical systems (MEMS/NEMS).[6] However, operating such devices requires complex circuitry to deliver the voltage via electrodes and electrical contacts. By contrast, photostriction (non-thermal strain generated by light) offers the possibility of remotely actuating ferroelectrics, without the need for electrodes, and with the faster operation afforded by the speed of light.

In addition, photostriction provides a window to explore light-matter interactions in ferroelectrics.[7,8] The mechanisms for generating photostriction in a material can be multiple: deformation potential interaction, i.e. electronic pressure on atomic bonds causing the lattice to expand or contract,[9] coupling between bulk photovoltaic effect (BPVE) and piezoelectricity,[10–13] depolarization field screening by photo-generated carriers,[14–16] suppression of long-range dipolar interaction by increased charge carrier density under illumination,[17–20] direct coupling to a phonon mode,[21] or indirect photon - electron - phonon interactions.[22] These mechanisms are not mutually exclusive, and happen across timescales ranging from sub-picoseconds to microseconds or even as slow as seconds. However, some of them are only available to polar materials. Often, the magnitude of photoinduced strain in thin films is significantly higher (0.41% for $BiFeO_3$ thin film) than what is observed in single crystals (14 x $10^{-4}$% for $BiFeO_3$ single crystal).[10,15,22] This is because the thickness of the films can be comparable to the penetration depth of photons, thus allowing a larger relative fraction of the material to participate in photostriction.

Thin films, however, are clamped to rigid substrates, which prevents the full development of photostrictive deformation. By contrast, freestanding films are unconstrained and have already demonstrated enhanced electromechanical properties.[23–25] Accordingly, photostrictive deformations are expected to be larger in freestanding thin films of ferroelectric materials than in clamped films. Recent advances in epitaxial lift-off techniques have enabled the production of fully-oriented freestanding thin films of complex oxides,[26,27] allowing the development of membrane resonators.[28,29] The stage is thus set for exploring optically-driven actuation in freestanding ferroelectric films.

In this work, we report opto-mechanical actuation of freestanding thin films of the prototype ferroelectric $BaTiO_3$ transferred onto perforated Silicon Nitride ($Si_3N_4$)//Silicon (Si) substrates. These freestanding films act as drum-like resonators when excited by a light beam, with resonant enhancement observed for oscillating illumination modulated around the mechanical resonance frequency of the nano-drums. Upon exposure to laser pulses, we observe deflections on the order of hundreds of nanometres in membranes that are only tens of nanometres thin. The negligible response of non-polar $SrTiO_3$ membranes,and the temperature dependence of photoactuation of $BaTiO_3$ membranes indicate that the out-of-plane deflection is induced by the coupling of light to ferroelectric polarization, and enhanced by the drumskin geometry of the freestanding membranes.

## 2 Results and Discussion

### 2.1 Fabrication and characterization of freestanding membranes

(001)-oriented $BaTiO_3$ thin films (10, 24 and 40 nm thick) and a $SrTiO_3$ thin film (20 nm) were respectively grown on (110)-oriented $GdScO_3$ and (001)-oriented $SrTiO_3$ substrates via pulsed laser deposition. Prior to the $BaTiO_3$ and $SrTiO_3$ growth, a sacrificial water-soluble layer of $Sr_3Al_2O_6$ (SAO) (8-10 nm) was deposited. This thin layer allows the epitaxial growth of $BaTiO_3$ and $SrTiO_3$, and dissolves when the sample is immersed in water, allowing the epitaxially-grown, coherently oriented top layers to be separated from the substrate, and transferred to a recipient substrate of choice (**Figure 1a**). To obtain freestanding films, we fabricated chips with through-holes having diameters ranging from 10 to 40 $\mu$m, on a 500 nm Silicon Nitride ($Si_3N_4$) membrane supported by a Silicon frame. The details of the microfabrication and schematic of the process is provided in Section 1 of Supporting Information. A polymer stamp process used here allowed us to transfer millimeter-sized regions of the films onto the holey $Si_3N_4$ membrane. Details of the substrate fabrication and transfer processes are provided in the Section 1 of the Supporting Information and Experimental Section. Despite the occurrence of cracks and delaminations, most of the membrane remains flat, including the areas over the substrate holes (**Figure 1b**). The topography of two typical membranes, acquired by Atomic Force Microscopy (AFM), are provided in Figure S1 of the Supporting Information, showing a smooth surface for both $BaTiO_3$ and $SrTiO_3$ with some residual curvature.

X-ray diffraction (XRD) (**Figure 1c**) illustrates that the crystallinity of the transferred membranes is well maintained upon releasing the thin film from substrate. The tetragonality of $BaTiO_3$ approximates that of bulk $BaTiO_3$,[30] as expected, given the strain relaxation after separation from the substrate. The same applies for $SrTiO_3$. From the XRD scans and reciprocal space maps (RSM) (Figure S2, Supporting Information) we only identify a single reflection from the membranes. This corresponds to an out-of-plane elongated tetragonal unit cell, suggesting that the single c-domain orientation of the epitaxial $BaTiO_3$ films is maintained even after the compressive strain from the substrate is removed. In other words, $BaTiO_3$ does not modify its microstructure or form in-plane domains upon declamping from the substrate.

Thereafter, we perform Piezoresponse Force Microscopy (PFM) that confirms the presence of ferroelectricity in $BaTiO_3$ membranes and discard it in $SrTiO_3$ (transferred onto Platinum coated Silicon) (**Figure 1d**). Electrically written domains in $BaTiO_3$ retained their phase contrast after a switching voltage is applied, while a weak contrast in $SrTiO_3$ disappears with time, consistent with charge injection rather than remnant polarization.

While the $BaTiO_3$ freestanding films are ferroelectric and the $SrTiO_3$ films are not, they are very similar mechanically. When pushed by an AFM tip in the centre of the membranes, both $BaTiO_3$ and $SrTiO_3$ get indented by almost equal amounts for a given force, indicating similar elastic moduli. The force vs. deflection curves are shown in Figure S3 of Supporting Information. In addition, $BaTiO_3$ and $SrTiO_3$ have similar optical bandgaps and absorbance profiles (Figure S4 of Supporting Information), making them akin in all relevant aspects (morphology, mechanical and optical properties). This allows us to single out the role of ferroelectricity in the photostrictive response of BaTiO$_3$, using SrTiO$_3$ as a paraelectric reference.

## 2.2 Optical actuation

The transferred membranes in the circularly-suspended regions have a geometry of drum-like resonators. To study their natural vibrations, we use a Laser Doppler Vibrometer (LDV).[31] A schematic of the experimental setup is shown in **Figure 2a** and the details of the measurement scheme are provided in the Experimental Section and the working principle of the LDV is explained in Section 2 of the Supporting Information. By focusing a 632.8 nm laser (0.2mW) and measuring the Doppler-shifted photon wavelength of the reflected light, it is possible to quantify the time-dependent deformation velocity. The amplitude of the deformation can be found by integrating the velocity over time. Using this technique, it is possible to identify several resonant oscillation modes on each drum resonator (Figure S5, Supporting Information), without the need of an external actuation, just by observing the thermomechanical noise of the membranes (Brownian exchange with the environment at each modal frequency, the sample being at a finite temperature).[32] From here on, we focus on the primary natural resonance mode of the resonators for further characterization. The frequency of the primary mode shows an inverse scaling with the diameter of the drums (**Figure 2b**), which is the signature of a resonator behaving like a thin membrane governed by the in-plane stress in the structure.[33] The corresponding equations governing the resonator behaviour are explained in the Section 3 of the Supporting Information.

To optically drive these membranes, we illuminate the sample with an analogically modulated semiconductor diode laser of central wavelength 405nm (3.06 eV) centred on the same spot as the LDV probing laser. This energy lies slightly below the bandgap of both BaTiO$_3$ and SrTiO$_3$ (Figure S4, Supporting Information). When sweeping the modulation frequency of the excitation laser around the natural resonance frequency, we record the optically-driven resonant response of the drums. BaTiO$_3$ and SrTiO$_3$ membranes of similar thickness and diameter show similar resonance frequencies of the fundamental mode **Figure 2c**, consistent with their similar mechanical properties. These responses are fitted to a Lorentzian function to calculate the resonance frequency ($f_0$) and quality factor (Q), the latter providing a measure of the energy dissipation of the resonators. The Q values, obtained by fitting the resonance peaks with a Lorentzian function shows a mean around 200±70 for both $BaTiO_3$ and $SrTiO_3$ membranes, comparable to those reported in other $SrTiO_3$ drum resonators.[28] Exceptionally, membranes with high Q values (1000 - 2000) were observed, the origin of which is still being studied. One example is provided in Figure S6 of the Supporting Information.

The resonators demonstrate a long linear regime of operation: when increasing the modulation amplitude of the laser, the amplitude of the resonant response of the drums increases proportionally. Normalizing this amplitude by Q, the responses of all $BaTiO_3$ resonators nearly overlap for each diameter value (**Figure 2d**). A similar scaling is observed for $SrTiO_3$ resonators (Figure 2d), but for the same diameter their oscillation amplitudes are three orders of magnitude lower compared to $BaTiO_3$. The different actuation efficiency of the ferroelectric resonators points to a role of polarization in the actuation mechanism.

To characterize the dynamics, we excite the membranes via laser pulses and measure the photoinduced deformation followed by its relaxation. When the laser is switched on, we observe a sudden change in the membrane's height (photoinduced deflection) (**Figure 3a**). The sign of the registered deflection is positive meaning that the membrane deforms out of the plane towards the detection path. When the laser is turned off, the deflection decays exponentially (for both $BaTiO_3$ and $SrTiO_3$), relaxing back to the initial state. Convolved with this exponential decay, we observe an oscillation with a frequency, corresponding to the fundamental resonant mode of the drum. These oscillations are most evident in measurements using short laser pulses (0.5 $\mu$s as shown in Figure 3a, and up to 10 $\mu$s not shown here), for which the initial deflections are lower and the pulse width is closer to the oscillation period of the resonant mode of the membrane (inset, Figure 3a). For longer pulse-widths (500 $\mu$s as shown here and 1 -10 $\mu$ s not shown here) we observe a large out-of-plane deflection followed by saturation (**Figure 3b**). Deflections as large as 250 nm are recorded for a 500 $\mu$s pulse, much larger than the thickness of the film (40 nm in this case). In contrast, the maximum deflection for the SrTiO$_3$ membrane (marked in blue) is $\sim 6$ nm, smaller than its thickness (20nm), despite the drum diameter being 30 $\mu$m for both and two orders of magnitude lower than SrTiO$_3$. The relaxation times, extracted from exponential fits to the decay, are independent of the input power and follow a linear dependence with the square of the diameter of the drums signifying surface losses not being the dominant damping mechanism (**Figure 3c**).[34] The maximum deflection scales linearly with the input laser power both BaTiO$_3$ and SrTiO$_3$ membranes (**Figure 3d**). These photoactuation amplitudes are considerably larger than those observed in epitaxial films (thanks to declamping) and faster than measured in polymeric or organic membranes.[35–37]

## 2.3 Crystallographic strain

To correlate the macroscopic deflection to crystallographic strain, we performed time-resolved X-ray microdiffraction in a pump-probe scheme, using the same laser source as the pump as in the vibrometry measurements (**Figure 4a**). We monitored the intensity around the (200) Bragg peak of BaTiO$_3$ as a function of the pump-probe delay, using a gateable X-ray detector synchronized with the laser modulation signal. Both the laser (spot of 24 × 35.5 $\mu$m$^2$) and the x-ray beam (1.9 × 1.9 $\mu$m$^2$) were focused on the same spot in the middle of a 40 $\mu$m diameter drum of a 24 nm thick membrane (more details of the measurement setup can be found in Experimental Section and Figure S7 Supporting Information). For a laser pulse width of 500 $\mu$s, we observed an in-plane lattice expansion of 0.047% (**Figure 4b**) when the laser is switched on. This lattice deformation develops with a rise-time of $\sim 60$ $\mu$s, and relaxes back to the initial state following a biexponential decay with two time constants: a fast one $\tau_1 = 33 \pm 4$ $\mu$s, matching the range of values extracted from LDV measurements and thus correlated with the photostriction mechanism behind the macroscopic drum deflection, and a secondary slower one $\tau_2 = 600 \pm 15$ $\mu$s (just over half a millisecond), which is small in magnitude and likely due to a drift or slow reconfiguration of the diffracting volume.

Due to geometrical constraints in the synchrotron hutch, the spot size of the excitation laser was bigger and hence the maximum peak power density was smaller (close to $3\,kW.cm^{-2}$). In order to compare results across the two measurement techniques and membranes, we interpolated the vibrometry deflection measurements down to the same power density as in the microdiffraction experiment. The extrapolated deflection would be 60 nm. With this value, the ratio between out-of-plane vertical displacement (d) and in-plane photogenerated expansion (diameter of the membrane increased by the measured 0.047% strain)(s) is d/s approximately 32. This is a measure of the "geometrical amplification", i.e. how much larger is the vertical displacement of the free-standing films (deflection) compared to what it would be if they were clamped to a rigid substrate (strain). The photoinduced in-plane expansion of the $BaTiO_3$ membrane, due to its clamping at the perimeter of the drum, can only be accommodated by a deflection of the structure out of the plane.(**Figure 4c**)

The ability of membranes to convert in-plane strain into out-of-plane deflection is somewhat reminiscent of the so-called "cymbal" actuators introduced by Uchino and Newnham in 1990's.[3, 38] Typical cymbal transducers consist of a flat piezoelectric capacitor clamped to a metallic dome-shaped endcap. When an electric field is applied to the piezoelectric, its in-plane deformation results in a change of curvature of the dome and concomitant perpendicular displacement of the endcap's apex. In comparison to cymbal piezo-actuators, however, the mechanism driving the membranes is different (light instead of voltage), and the geometry is simpler: while in cymbals the strain is generated in one element (piezoelectric capacitor) and converted into deflection by another (metallic dome), in the nano-drums the strain and the deflection happen in the same element: an electrode-free and end-cap-free photostrictive membrane.

## 2.4   Electronic origin of photostriction

Having quantified the magnitude of the photostriction, the next question is what is the light-matter interaction mechanism that causes it. The following ferroelectric - specific photostriction mechanisms are considered: combined action of bulk photovoltaic effect and inverse piezoelectric effect, laser-induced local heating causing thermal expansion and/or pyroelec-tricity, or screening of long range dipolar order by photocarriers.

The bulk photovoltaic effect (BPVE) generates a photovoltage due to carrier separation in a non-centrosymmetric lattice, which combined with the inverse piezoelectric effect modifies the strain state. In perovskite ferroelectrics, this effect results in an expansion of the lattice parameter along the polar axis.[10,39] Therefore, in $BaTiO_3$ membranes it would manifest itself as an increased tetragonality, i.e., an increased out-of-plane ('c') lattice parameter and decreased in-plane ('a') lattice parameter. However, X-ray micro-diffraction (Figure 4b) indicates the opposite: the in-plane lattice parameter expands upon illumination. We therefore discard BPVE as the principal source of strain.

Contrary to the BPVE, however, photo-induced heating should reduce tetragonality and therefore produce an opposite behaviour of the lattice parameters: contraction of the polar axis and elongation of the non-polar axis.[40] In order to determine the lattice response to pure thermal effects, we measured temperature dependent X-ray diffraction scans and RSM on a 40nm $BaTiO_3$ membrane (Figure S8 of Supporting Information), from which we extract two pieces of information: (i) the Curie temperature of the membrane is around 200ºC, as indicated by the dip in the out-of-plane parameter and (ii) while the out-of-plane lattice parameter decreases towards $T_c$ and then increases above, the in-plane lattice parameter grows at a constant rate ($2.5 \times 10^{-5}$ $K^{-1}$) throughout the temperature range. This value of the expansion coefficient is similar to the expansion coefficient of cubic $SrTiO_3$.[41] Therefore, if thermal expansion was the ruling mechanism for the photoinduced deflection, we would expect similar behaviour in both $BaTiO_3$ and $SrTiO_3$ membranes, which is not the case.

Nevertheless, opto-thermal heat can have an additional effect on ferroelectrics via pyroelectricity, which is the coupling between polarization and temperature. It is hypothetically possible that a photo-induced temperature change causes a change in polarization and therefore a pyroelectric voltage,[42] which in turn, could modify the strain via inverse piezoelectricity. This would be somewhat analogous to the coupling between the bulk photovoltaic effect and inverse piezoelectricity, but with the interaction being mediated not by a photovoltage but by a pyrovoltage. However, the pyroelectric coefficient is proportional to the rate of change of polarization as a function of temperature and it is therefore not constant; it increases rapidly as the Curie temperature ($T_c$) is approached, and it vanishes above $T_c$.[43] If a pyroelectric voltage was the driving mechanism for photostriction, and therefore the amplitude of the photoresponse would mirror the temperature dependence of the pyroelectric coefficient, increasing towards $T_c$.

In order to verify the link between polarization and photostriction, we measured the temperature dependence of photoactuation, using digital holographic microscopy (DHM); an interferometric microscopy method with nanometric vertical resolution (**Figure 5a**).[44] We observe that the magnitude of photo-deflection is maximal at room temperature and decreases with increasing temperature (**Figure 5b**). This magnitude as a function of temperature, when compared with the tetragonality as a function of temperature for the same membrane, follows the same trend as expected for the evolution of polar order parameter(**Figure 5c**),[45, 46] proving that photoactuation in $BaTiO_3$ membranes is a direct consequence of the light interacting with the polarization. In other words, the photostrictive mechanism is proportional to polarization, not to its rate of change, and hence not driven by pyroelectricity.

That leaves us only the theoretical mechanism proposed by Paillard et. al.: the appearance of photocarriers screens the polarization, thus shortening the out-of-plane lattice parameter and therefore tetragonality.[18,19] Suppression of polarization by photocarriers has been observed in $BaTiO_3$ epitaxial films and single crystals before,[20,47] and is consistent with the observed elongation of the in-plane parameter in the micro-diffraction experiment (**Figure 5d**).

There is no predictive model for the magnitude of photoscreening-mediated strain in the tetragonal phase of $BaTiO_3$, but there are first-principle calculations for the tetragonal phase of $PbTiO_3$ and for the rhombohedral ground state of $BaTiO_3$.[12] The predicted change in strain ($\epsilon$) caused by photoinduced carrier density (n) ($\delta\epsilon/\delta n$) is in the range 5 to 20 × $10^{20}$ $cm^{-3}$.[12] Taking this range of $\delta\epsilon/\delta n$ , for the photostriction observed in the diffraction experiment (0.047%), the photo-generated carrier density should be of the order, 0.8 to 20 × $10^{20}$ $cm^{-3}$. Given the optical absorbance of our films (0.094 au) and the reported quantum efficiency of BaTiO$_3$ (between 2.1 × 10$^{-4}$ and 1 × 10$^{-2}$),[48, 49] the photo-carrier density generated per light pulse in our sample is of the order of 0.3 to 15 × 10$^{20}$ cm$^{-3}$, which is in remarkably good agreement with the theoretical estimate. Further details of the calculation are provided in Section 4 of the Supporting Information.

Finally, we attribute the marginal photostrictive effects observed in SrTiO$_3$ membranes to a combination of thermal effects and possibly other mechanisms as discussed in the introduction, which do no require any symmetry constraints such as the deformation pòtential interaction.[9] The difference between BaTiO$_3$ and SrTiO$_3$ is that, in the ferroelectric, the change in inter-atomic distances is anisotropic and exacerbated by the screening of the polarization-dependent spontaneous strain (Paillard's model),[12] whereas in the later the changes will be smaller as there is no polarization to screen, and thus the change in bond distances is small and orthotropic. Thus, the photocarrier excitation causes a larger response in the ferroelectric and a small one in non-ferroelectric one, as observed in our study.

## 3 Conclusion

In contrast to epitaxial films clamped to rigid substrates,[16, 39] freestanding ferroelectric membranes have a vertical degree of freedom for motion that allows them to accommodate large deflections as a result of a photostriction amplified by the suspended geometry. The measured in-plane lattice expansion upon illumination and the temperature dependence of photoactuation, both point to the photostrictive mechanism being polarization suppression by photoexcited carriers.

Beyond shedding light on the dominant mechanism for photostriction in BaTiO$_3$ thin films, the results are an exciting development for electrode-free actuators. Our observations also reveal a wide range of control knobs for the photoactuation via external parameters (laser power, frequency, pulse width, temperature) or sample design parameters (suspended area, film thickness). This wide range of tunability, together with the potential of freestanding membranes to be integrated on semiconductor or flexible devices, or combined on multifunc-tional heterostructures,[50] or to adopt different geometries such as wrinkles,[24] tubes,[51] or ribbons,[52] makes our study very appealing for designing photoresponsive actuators at the nanoscale.[37] Furthermore, we foresee large room for improvement in the efficiency of the photoactuation using ferroelectrics with lower bandgap ($BiFeO_3$) or utilizing materials that display giant photostriction effects.[53]

# 4   Experimental Section

***Epitaxy and Fabrication of Single-Crystal Membranes***:   All films were grown using pulsed-laser deposition using a KrF excimer laser (248 nm, COMPex 102, Lambda Physik). All $BaTiO_3$ thin films studied herein were grown on $GdScO_3$ (110) substrates (CrysTec GmbH) at a temperature of 750ºC and oxygen pressure of 100mTorr while the underlying $Sr_3Al_2O_6$ layer was grown at pressure of 1mTorr. The laser fluence was 1.6 J.cm$^{-2}$. All layers were grown from ceramic stoichiometric targets (Praxair) of the same chemistry as the films. Following growth, samples were cooled to room temperature at 10ºC min$^{-1}$ in a static oxygen pressure of 10 Torr. On the other hand, $SrTiO_3$ films were grown on $SrTiO_3$ (100) substrates, keeping the rest of the conditions same. To release the films from the substrates, the epitaxial stack was placed on a poly-dimethylsiloxane (PDMS) stamp (Sylgard184 from Dow Corning). The samples were then placed in Milli-Q water and removed from it when the $Sr_3Al_2O_6$ was fully dissolved, as ascertained by visual inspection. The freestanding films attached to the PDMS stamp were then transferred to $Si_3N_4$ membranes supported on a Silicon frame. Additional details of the transfer process and the fabrication of the Silicon frame with the membrane are provided in the Section 1 of Supporting Information. Both $BaTiO_3$ and $SrTiO_3$ thin films undergo no thermal processing except during the transfer process (after the release, when the sacrificial layer has dissolved) on the Silicon Nitride membrane when the stack is heated up to 70ºC to facilitate the adhesion between the thin film and the Silicon Nitride membrane and during the temperature dependent measurements (Vibrometry, XRD), when the samples are heated up to 450ºC.

***Structural, Surface and Electrical Characterization***:   X-ray diffraction studies were performed on a Panalytical X'pert Pro diffractometer (Copper K$_{\alpha 1}$), using a parabolic mirror and a (220)Ge 2-bounce monochromator on the incident beam side and a PIXcel position-sensitive detector. This includes the $2\theta$-$\omega$ scans and reciprocal space maps. The offset angle $\omega$ for the membrane samples was obtained by averaging the peak $\omega$ value of the same reflection acquired at azimuthal angles of $\Phi = 0°$ and $\Phi = 180°$. High-temperature diffraction experiments were carried out in ambient atmosphere using a Anton Paar DHS1100 domed hot stage.

All surface topography and piezoresponse characterization was performed by MFP-3D Asylum AFM (Asylum Research -Oxford Instruments). For PFM measurements, PPP-EFM tips (Nano sensors; Schaffhausen, Switzerland) were used while for topography and indentation of membranes OPUS 240AC-PP soft tips (k = 2 N.m$^{-1}$) were used.

***Optical absorbance***:   Absorbance of BaTiO$_3$ and SrTiO$_3$ membranes were measured using a Cary4000 UV-Vis spectrophotometer. Thin films of BaTiO$_3$ and SrTiO$_3$ were transferred on poly-dimethylsiloxane (PDMS) pieces and were measured as such while the absorbance spectra from PDMS was used for background correction. The measured data contained absorbance in absorbance units. A Tauc plot was constructed from absorbance data with a direct bandgap exponent. The tangent to the linear part of the plot was extrapolated to estimate the direct optical bandgap.

***Laser Doppler Vibrometry***:   The optomechanical response of the membranes was characterized by a modular Laser Doppler Vibrometer (LDV) control unit from Polytec GmBH, that consists of He-Ne laser (probe laser) for detection and decoders which use analog to digital converters for processing the output signal. This signal was fed to a Lock-In Amplifier (UHFLIA(600 MHz), Zurich Instruments) through which data was acquired. The working principle of a Doppler vibrometer is explained in Section 2 of the Supporting Information. A graphical representation of the measurement scheme is depicted in Figure 2 of main text. The thermomechanical noise of the membranes were recorded using the same scheme through the lock-in. For exciting the membranes, a 405nm (centre wavelength) fibre pigtailed semiconductor diode laser (Cobolt 06-MLD) was coincided with a part of the detection path through the objective, that focuses both the probe laser and Cobolt laser to spot sizes of 5.45 $\mu$m and 8 $\mu$m respectively. The incident laser had a polarization parallel to the surface of the membranes (orthogonal to the out-of-plane polarization of BaTiO$_3$). The Cobolt laser controller was connected to the lock-in amplifier for analog or digital modulation. The samples were placed in a vacuum chamber at a pressure of 0.01 mbar, to minimize viscous losses with the surrounding atmosphere.

The resonance measurements were done by sweeping in a frequency range (by analog modulation of the Cobolt laser) and detecting the output in voltage units. This can be converted to displacement units by a calibration factor of 50 nm.V$^{-1}$. In analog modulation mode, a fixed offset of 700 mV was applied to keep the laser on , while an additional sinusoid of 0.1 mV, 1 mV, 10 mV and 100 mV to give the input excitation for actuating the membranes. The average power density for 700 mV offset was around 60 W.cm$^{-2}$.

For measuring the photoinduced effect and its dynamics, the pump laser was modulated digitally through a pulse signal generated from the arbitrary waveform generator of the lock-in amplifier. This generated pulse was verified through a balanced 80MHz NewFocus photodetector (Model 1807 - FC) read in an oscilloscope. The generated pulses had rise time of 10 ns. The maximum emitted peak power was controlled from the laser controller. The actual peak power density on the sample corresponding to 5 different peak powers used in the experiment are, 0.8, 2, 4, 6 and 8 kW.cm$^{-2}$.

***Digital Holographic Microscopy***:   A commercial Digital Holographic Microscope (DHM), R-1000, from Lyncee Tec, was adapted to do stroboscopic measurements with optical excitation. The DHM, essentially an interferometer, uses the phase-shifted reflected light from the sample surface to interfere with a reference beam, giving a phase and amplitude information. The phase information is then internally converted to topographic data. The same COBOLT 405nm modulated laser, as mentioned above, was used as the excitation source. An internal TTL signal is used to trigger the laser. The data acquisition was done through KOALA software package and analyzed in the MEMS analysis tool. A peak power density of 4 kW.cm$^{-2}$ was used for the holographic measurements.

***Time-resolved X-ray micro-diffraction***:   Time-resolved X-ray diffraction measurements were performed in a pump-probe mode at the CRISTAL beamline of SOLEIL Synchrotron. The incoming beam with an energy of 8.5 keV was focused at the sample plane with a spot size of approximately 1.9 x 1.9 $\mu$m (FWHM) using Fresnel zone plates. The sample was placed in diffraction condition for the (200) reflection. The details of the diffraction conditions and geometry are provided in the Supporting Information. At the sample plane the pump laser was focused with an oblique angle, because of the spatial limitations of the setup. The gated XPAD3.2 detector and the pump laser were synchronized with with an externally-triggered delay generator,[54] allowing the detector to be switched on for specific user-defined time delays with respect to the t= 0 of the pump laser. The time resolution of the experiment was 300 ns, set by the duration of the gate applied to the XPAD3.2 detector. The projected peak power density of the excitation laser was 3 kW.cm$^{-2}$.

**Supporting Information**: Supporting Information is available from the Wiley Online Library or from the author.

**Acknowledgements**: The authors acknowledge Prof. Josep Fontcuberta for insightful discussions, Dr. Javier Saiz for help with absorbance measurements, and Dr. Emigdio Chavez for discussions about experimental setup. The authors also acknowledge SOLEIL for provision of synchrotron radiation at CRISTAL beamline (proposal number 20210988). S.G acknowledges the PREBIST Cofund grant. This project has received funding from the European Union's Horizon 2020 research and innovation programme under the Marie Skłodowska-Curie grant agreement No. 754558. D.P acknowledges funding from 'la Caixa' Foundation fellowship (ID 100010434). The authors acknowledge the Spanish Ministry of Industry, Economy and Competitiveness (MINECO) through Grants No. PID2019-108573GB-C21 and PID2019-109931GB-I00. The ICN2 is funded by the CERCA programme / Generalitat de Catalunya and by the Severo Ochoa Centres of Excellence Programme, funded by the Spanish Research Agency (AEI, CEX2021-001214-S).

# References

[1] Y. Lee, J. Park, S. Cho, Y.-E. Shin, H. Lee, J. Kim, J. Myoung, S. Cho, S. Kang, C. Baig, H. Ko, *ACS Nano* **2018**, *12*, 4, 4045–4054.

[2] P. Poosanaas, K. Tonooka, K. Uchino, *Mechatronics* **2000**, *10*, 4-5, 467–487.

[3] Y. Sugawara, K. Onitsuka, S. Yoshikawa, Q. Xu, R. E. Newnham, K. Uchino, *Journal of the American Ceramic Society* **1992**, *75*, 4, 996–998.

[4] N. Izyumskaya, Y. Alivov, H. Morkoc, *Critical Reviews in Solid State and Materials Sciences* **2009**, *34*, 3-4, 89–179.

[5] C. Bowen, H. Kim, P. Weaver, S. Dunn, *Energy & Environmental Science* **2014**, *7*, 1, 25–44.

[6] S. Trolier-McKinstry, P. Muralt, *Journal of Electroceramics* **2004**, *12*, 1, 7–17.

[7] C. Chen, Z. Yi, *Advanced Functional Materials* **2021**, *31*, 22, 2010706.

[8] B. Kundys, *Applied Physics Reviews* **2015**, *2*, 1, 011301.

[9] T. Figielski, *physica status solidi (b)* **1961**, *1*, 4, 306–316.

[10] B. Kundys, M. Viret, D. Colson, D. O. Kundys, *Nature Materials* **2010**, *9*, 10, 803–805.

[11] C. Paillard, B. Xu, B. Dkhil, G. Geneste, L. Bellaiche, *Physical Review Letters* **2016**, *116*, 24, 247401.

[12] C. Paillard, S. Prosandeev, L. Bellaiche, *Physical Review B* **2017**, *96*, 4, 045205.

[13] D. Schick, M. Herzog, H. Wen, P. Chen, C. Adamo, P. Gaal, D. G. Schlom, P. G. Evans, Y. Li, M. Bargheer, *Physical Review Letters* **2014**, *112*, 9, 097602.

[14] D. Daranciang, M. J. Highland, H. Wen, S. M. Young, N. C. Brandt, H. Y. Hwang, M. Vattilana, M. Nicoul, F. Quirin, J. Goodfellow, T. Qi, I. Grinberg, D. M. Fritz, M. Cammarata, D. Zhu, H. T. Lemke, D. A. Walko, E. M. Dufresne, Y. Li, J. Larsson,

D. A. Reis, K. Sokolowski-Tinten, K. A. Nelson, A. M. Rappe, P. H. Fuoss, G. B. Stephenson, A. M. Lindenberg, *Physical Review Letters* **2012**, *108*, 8, 087601.

[15] H. Wen, P. Chen, M. P. Cosgriff, D. A. Walko, J. H. Lee, C. Adamo, R. D. Schaller, J. F. Ihlefeld, E. M. Dufresne, D. G. Schlom, P. G. Evans, J. W. Freeland, L. Yuelin, *Physical Review Letters* **2013**, *110*, 3, 037601.

[16] Y. Li, H. Wen, P. Chen, M. P. Cosgriff, D. Walko, J. H. Lee, C. Adamo, R. Schaller, C. Rowland, C. Schlepuetz, E. Dufresne, Q. Zhang, C. Giles, D. Schlom, J. Freeland, P. Evans, *MRS Online Proceedings Library* **2013**, *1528*, 1, 1–6.

[17] V. M. Fridkin, *Photoferroelectrics*, volume 9, Springer Science & Business Media, **2012**.

[18] C. Paillard, E. Torun, L. Wirtz, J. Íñiguez, L. Bellaiche, *Physical Review Letters* **2019**, *123*, 8, 087601.

[19] Y. Wang, X. Liu, J. D. Burton, S. S. Jaswal, E. Y. Tsymbal, *Physical Review letters* **2012**, *109*, 24, 247601.

[20] X. Long, H. Tan, F. Sánchez, I. Fina, J. Fontcuberta, *Nature Communications* **2021**, *12*, 1, 1–9.

[21] X. Li, T. Qiu, J. Zhang, E. Baldini, J. Lu, A. M. Rappe, K. A. Nelson, *Science* **2019**, *364*, 6445, 1079–1082.

[22] T.-C. Wei, H.-P. Wang, H.-J. Liu, D.-S. Tsai, J.-J. Ke, C.-L. Wu, Y.-P. Yin, Q. Zhan, G.-R. Lin, Y.-H. Chu, J.-H. He, *Nature Communications* **2017**, *8*, 1, 1–8.

[23] G. Dong, S. Li, M. Yao, Z. Zhou, Y.-Q. Zhang, X. Han, Z. Luo, J. Yao, B. Peng, Z. Hu, H. Huang, T. Jia, J. Li, W. Ren, Z.-G. Ye, X. Ding, J. Sun, C.-W. Nan, L.-Q. Chen, J. Li, M. Liu, *Science* **2019**, *366*, 6464, 475–479.

[24] G. Dong, S. Li, T. Li, H. Wu, T. Nan, X. Wang, H. Liu, Y. Cheng, Y. Zhou, W. Qu, Y. Zhao, B. Peng, Z. Wang, Z. Hu, Z. Luo, W. Ren, S. J. Pennycook, J. Li, J. Sun,

Z.-G. Ye, Z. Jiang, Z. Zhou, X. Ding, T. Min, M. Liu, *Advanced Materials* **2020**, *32*, 50, 2004477.

[25] H. Elangovan, M. Barzilay, S. Seremi, N. Cohen, Y. Jiang, L. W. Martin, Y. Ivry, *ACS Nano* **2020**, *14*, 4, 5053–5060.

[26] D. Pesquera, A. Fernández, E. Khestanova, L. W. Martin, *Journal of Physics: Condensed Matter* **2022**, *34*, 38, 383001.

[27] F. M. Chiabrera, S. Yun, Y. Li, R. T. Dahm, H. Zhang, C. K. Kirchert, D. V. Christensen, F. Trier, T. S. Jespersen, N. Pryds, *Annalen der Physik* **2022**, *534*, 9, 2200084.

[28] D. Davidovikj, D. J. Groenendijk, A. M. R. Monteiro, A. Dijkhoff, D. Afanasiev, M. Šiškins, M. Lee, Y. Huang, E. van Heumen, H. van der Zant, A. D. Caviglia, P. G. Steeneken, *Communications Physics* **2020**, *3*, 1, 1–6.

[29] M. Lee, J. R. Renshof, K. J. van Zeggeren, M. J. Houmes, E. Lesne, M. Šiškins, T. C. van Thiel, R. H. Guis, M. R. van Blankenstein, G. J. Verbiest, A. D. Caviglia, H. S. J. van der Zant, P. G. Steeneken, *Advanced Materials* **2022**, 2204630.

[30] G. Kwei, A. Lawson, S. Billinge, S. Cheong, *The Journal of Physical Chemistry* **1993**, *97*, 10, 2368–2377.

[31] C. Rembe, G. Siegmund, H. Steger, M. Wörtge, In *Optical inspection of microsystems*, 297–347. CRC Press, **2019**.

[32] S. Schmid, L. G. Villanueva, M. L. Roukes, *Fundamentals of nanomechanical resonators*, volume 49, Springer, **2016**.

[33] T. Wah, *Journal of the Acoustical Society of America* **1962**, *34*, 3, 275–281.

[34] L. G. Villanueva, S. Schmid, *Physical Review Letters* **2014**, *113*, 22, 227201.

[35] T. Vasileiadis, T. Marchesi D'Alvise, C.-M. Saak, M. Pochylski, S. Harvey, C. V. Synatschke, J. Gapinski, G. Fytas, E. H. Backus, T. Weil, B. Graczykowski, *Nano Letters* **2021**, *22*, 2, 578–585.

[36] W. H. Liew, Y. Chen, M. Alexe, K. Yao, *Small* **2022**, *18*, 7, 2106275.

[37] J. Li, X. Zhou, Z. Liu, *Advanced Optical Materials* **2020**, *8*, 18, 2000886.

[38] A. Dogan, J. F. Fernandez, K. Uchino, R. E. Newnham, In *ISAF'96. Proceedings of the Tenth IEEE International Symposium on Applications of Ferroelectrics*, volume 1. IEEE, **1996** 213–216.

[39] S. Matzen, L. Guillemot, T. Maroutian, S. K. Patel, H. Wen, A. D. DiChiara, G. Agnus, O. G. Shpyrko, E. E. Fullerton, D. Ravelosona, P. Lecoeur, R. Kukreja, *Advanced Electronic Materials* **2019**, *5*, 6, 1800709.

[40] P. W. Forsbergh Jr, *Physical Review* **1949**, *76*, 8, 1187.

[41] D. de Ligny, P. Richet, *Physical Review B* **1996**, *53*, 6, 3013.

[42] L. Wang, F. Zhang, C. Chen, X. He, M. A. Boda, K. Yao, Z. Yi, *Nano Energy* **2024**, *119*, 109081.

[43] T. Perls, T. Diesel, W. Dobrov, *Journal of Applied Physics* **1958**, *29*, 9, 1297–1302.

[44] U. K. Bhaskar, N. Banerjee, A. Abdollahi, Z. Wang, D. G. Schlom, G. Rijnders, G. Catalan, *Nature Nanotechnology* **2016**, *11*, 3, 263–266.

[45] H. F. Kay, P. Vousden, *The London, Edinburgh, and Dublin Philosophical Magazine and Journal of Science* **1949**, *40*, 309, 1019–1040.

[46] K. J. Choi, M. Biegalski, Y. Li, A. Sharan, J. Schubert, R. Uecker, P. Reiche, Y. Chen, X. Pan, V. Gopalan, L. Chen, D. G. Schlom, C. B. Eom, *Science* **2004**, *306*, 5698, 1005–1009.

[47] J. Wang, B. Vilquin, N. Barrett, *Applied Physics Letters* **2012**, *101*, 9, 092902.

[48] G. Unal, Measurements of the microscopic properties contributing to photorefraction in barium titanate, Technical report, Naval Postgraduate School, Monterey, **1986**.

[49] K. H. Yoon, K. S. Chung, *Journal of Applied Physics* **1992**, *72*, 12, 5743–5749.

[50] D. Pesquera, E. Khestanova, M. Ghidini, S. Zhang, A. P. Rooney, F. Maccherozzi, P. Riego, S. Farokhipoor, J. Kim, X. Moya, M. Vickers, N. A. Stelmashenko, S. Haigh, S. S. Dhesi, N. D. Mathur, *Nature Communications* **2020**, *11*, 1, 3190.

[51] Y. Li, E. Zatterin, M. Conroy, A. Pylypets, F. Borodavka, A. Björling, D. J. Groenendijk, E. Lesne, A. J. Clancy, M. Hadjimichael, D. Kepaptsoglou, Q. Ramasse, A. D. Caviglia, J. Hlinka, U. Bangert, S. J. Leake, P. Zubko, *Advanced Materials* **2022**, *34*, 15, 2106826.

[52] D. Kim, M. Kim, S. Reidt, H. Han, A. Baghizadeh, P. Zeng, H. Choi, J. Puigmartí-Luis, M. Trassin, B. J. Nelson, X.-Z. Chen, S. Pané, *Nature Communications* **2023**, *14*, 1, 750.

[53] Y. Zhou, L. You, S. Wang, Z. Ku, H. Fan, D. Schmidt, A. Rusydi, L. Chang, L. Wang, P. Ren, L. Chen, G. Yuan, L. Chen, J. Wang, *Nature Communications* **2016**, *7*, 1, 11193.

[54] K. Medjoubi, S. Hustache, F. Picca, J.-F. Bérar, N. Boudet, F. Bompard, P. Breugnon, J.-C. Clémens, A. Dawiec, P. Delpierre, B. Dinkespiler, S. Godiot, J.-P. Logier, M. Menouni, C. Morel, M. Nicolas, P. Pangaud, E. Vigeolas, *Journal of Instrumentation* **2011**, *6*, 01, C01080.

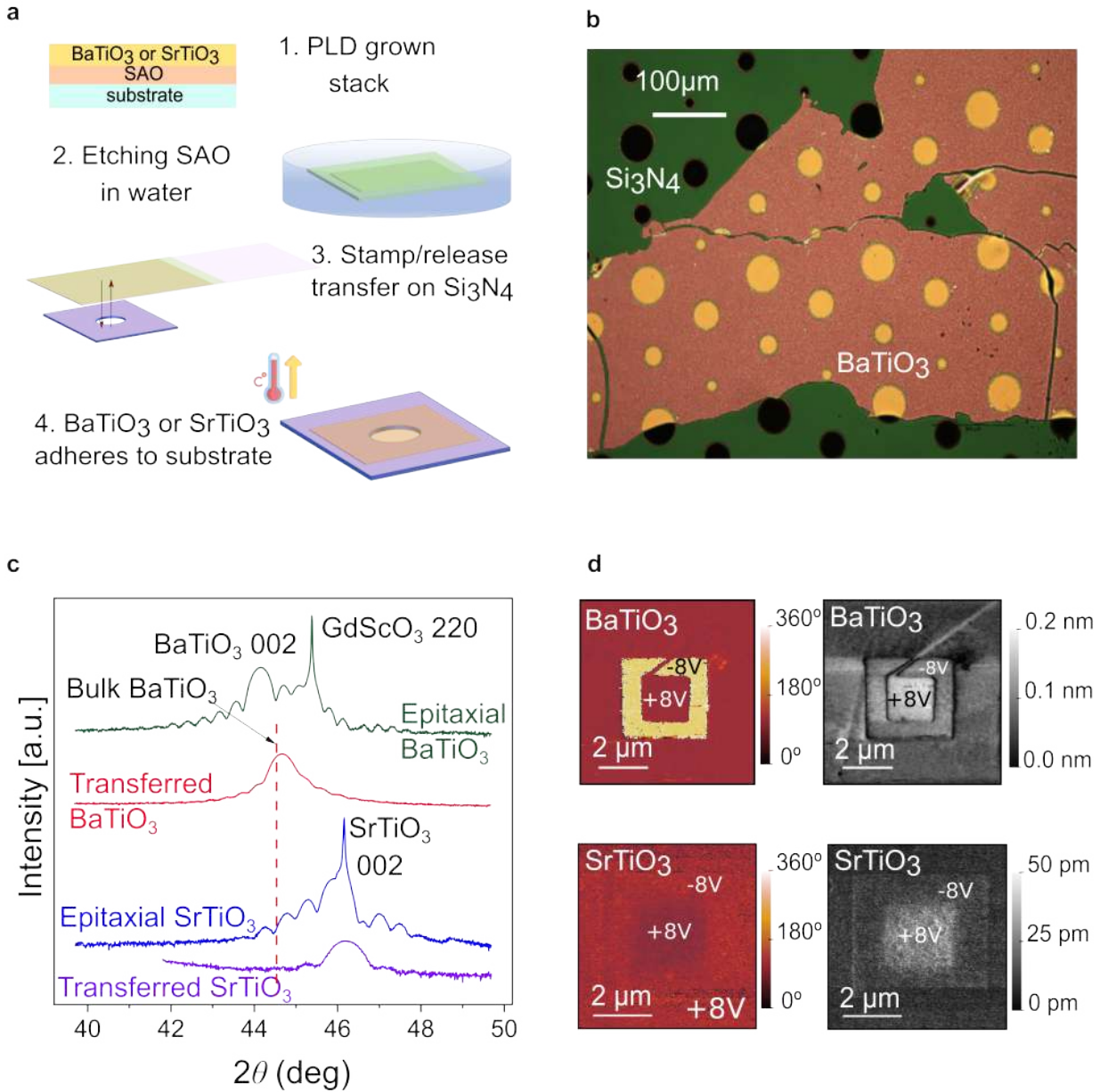

Figure 1: a. Schematic of sacrificial layer dissolution and thin film transfer process. b. Optical microscope image of a $BaTiO_3$ thin film suspended over holey $Si_3N_4$ membrane c. X-ray diffractograms of epitaxial and released $BaTiO_3$ and $SrTiO_3$ thin films. d. Piezoresponse force microscopy of $BaTiO_3$ and $SrTiO_3$ membranes transferred on Platinum-coated Silicon substrate (phase on left, amplitude on right), after "writing" a square pattern with a DC voltage. The $BaTiO_3$ films show 180 degree phase contrast when applying negative/positive voltage to the tip (outer/inner square), and amplitude contrast at the boundaries, consistent with switching of ferroelectric polarization and electrical writing of domains. While the $SrTiO_3$ film shows negligible phase contrast and weak amplitude contrast consistent with charge injection and no ferroelectricity.

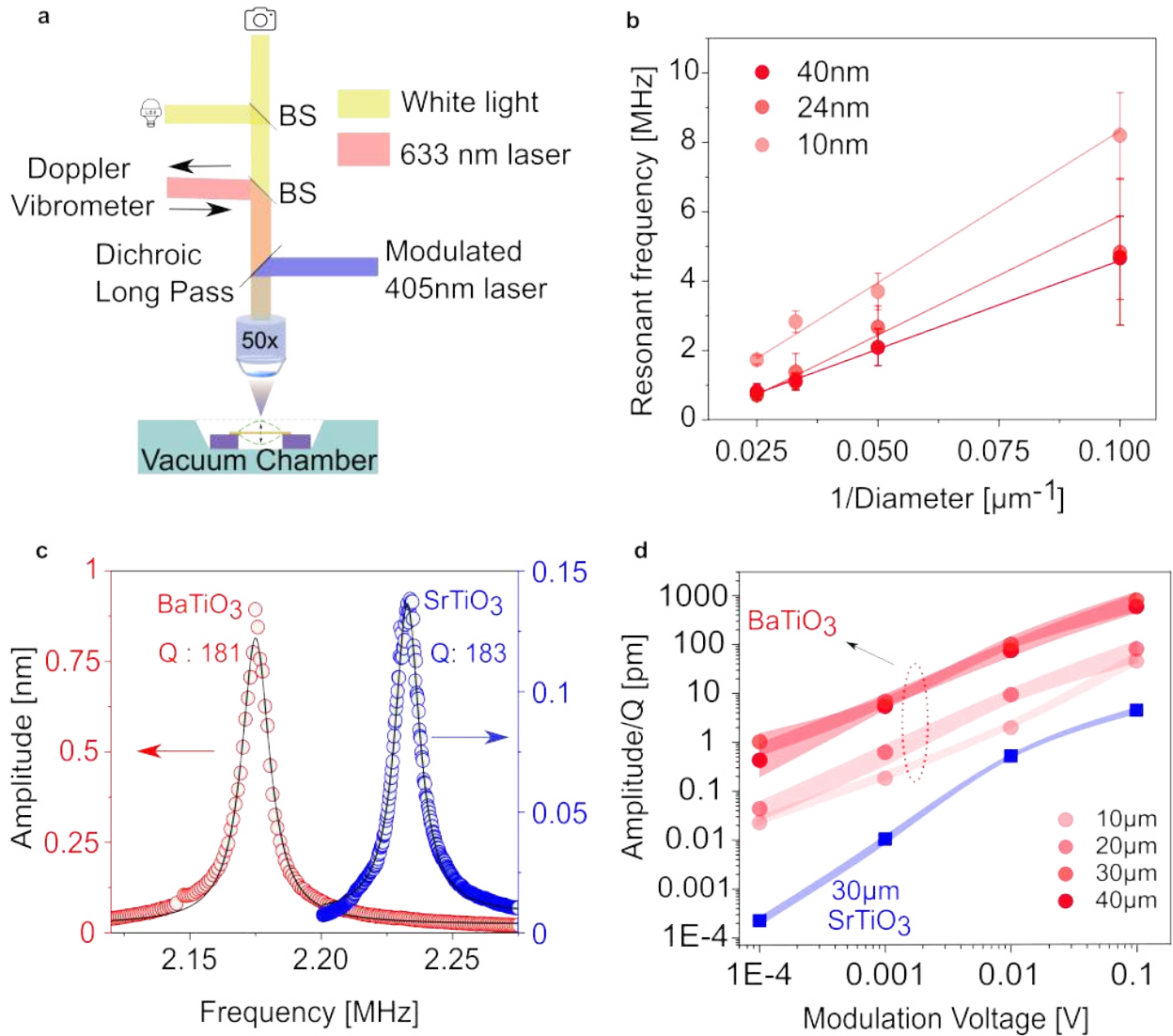

Figure 2: a. Schematic of the Doppler Vibrometry setup. The sample is imaged through a Quartz window of the vacuum chamber by a back-reflection path and a 50x objective coupled to a white light source and camera. The excitation laser and probe laser is coupled to the same path by a dichroic mirror and a beamsplitter (BS) respectively. b. Resonance frequency of the BaTiO₃ membranes scale linearly with the inverse of diameter, which is typical of membranes governed by residual strain rather than flexural rigidity c. Resonance curves of a BaTiO₃ (24 nm) and SrTiO₃ (20 nm) membrane of 30 micron diameter. d. Comparison of amplitude at resonance for BaTiO₃ (24 nm) and SrTiO₃ (20 nm) membranes of different diameters normalized by the Q factor. Shades denote standard deviation.

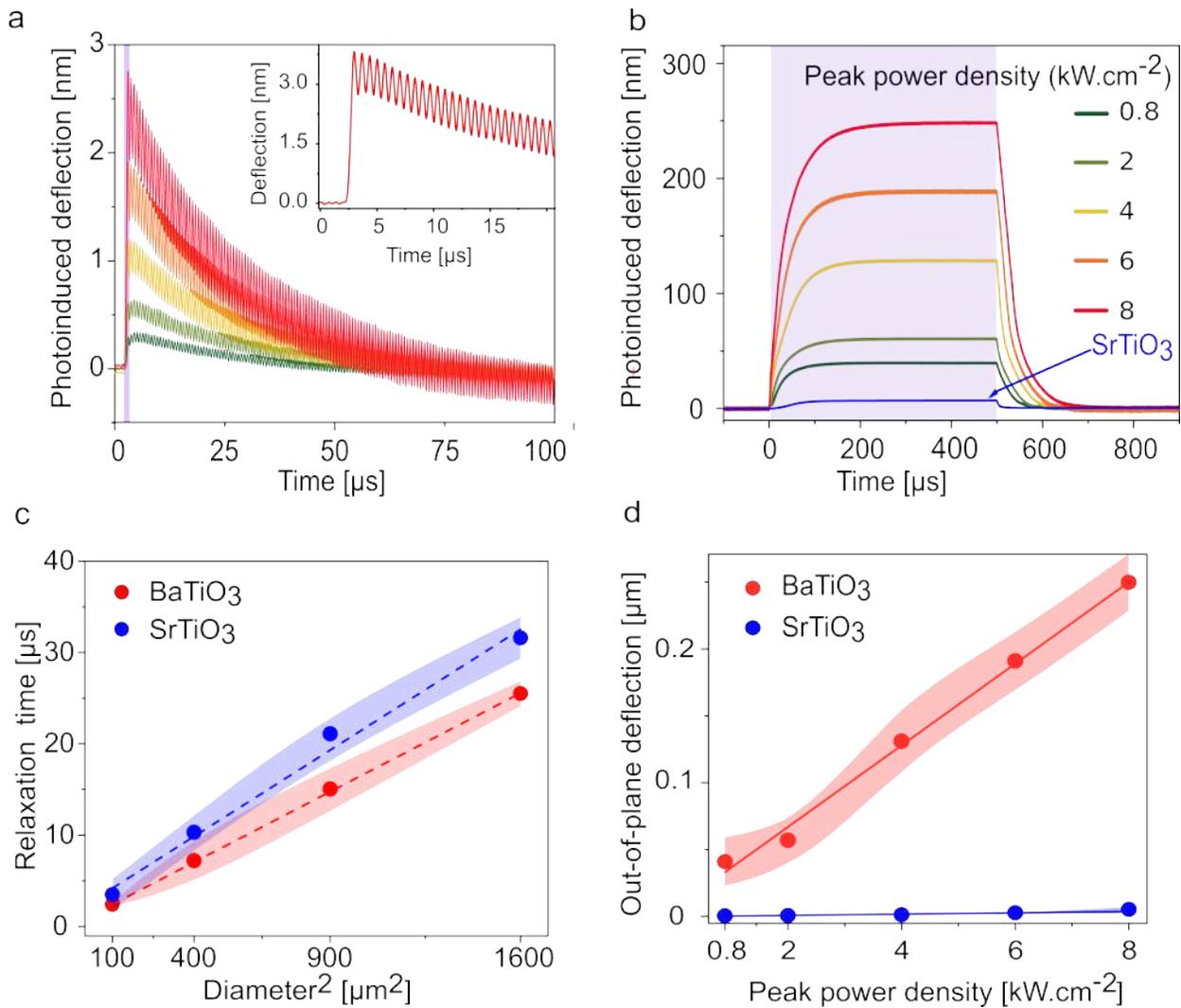

Figure 3: Pulsed excitation of $BaTiO_3$ membrane (30$\mu$m) with increasing peak power density of the laser (shaded area in purple marks the pulse width). a. Excitation by short pulses (0.5$\mu$s) creates small out-of-plane deflection overlaid with oscillations (inset) of the membrane vibrating at its natural resonant frequency. b. At longer pulse width (500$\mu$s), the out-of-plane deflection increases and saturates displaying large out-of-plane deflection. In contrast, $SrTiO_3$ membranes (blue) under same excitation respond two orders of magnitude lower. For $SrTiO_3$, a peak power density of 8 kW.cm$^{-2}$ is used c. Relaxation time of the out-of-plane deflection linearly proportional to the square of the diameter. This is obtained by fitting an exponential to the decay profile from measurements, shown in panel 'b'. Relaxation times are independent of laser power density d. Comparison of out-of-plane deflection for 500$\mu$s pulses for $BaTiO_3$ membranes and $SrTiO_3$ membranes. Shaded area in data denotes standard deviation.

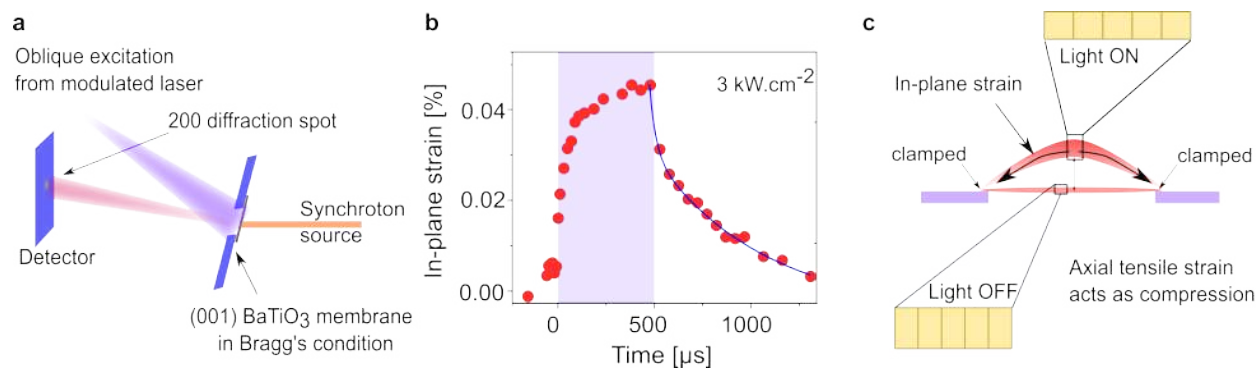

Figure 4: a. Schematic of the time-resolved X-ray microdiffraction setup. b. Relative change in in-plane strain calculated from transient of the (200) diffraction peak. The relaxation is fitted with two exponential decay function (blue). c. Schematic of the process resulting in actuation of the membranes. Upon shining light, polarization screening reduces the tetragonality of unit cell while elongating in-plane. The membrane being clamped on edges, the in-plane expansion results in the out-of-plane deflection.

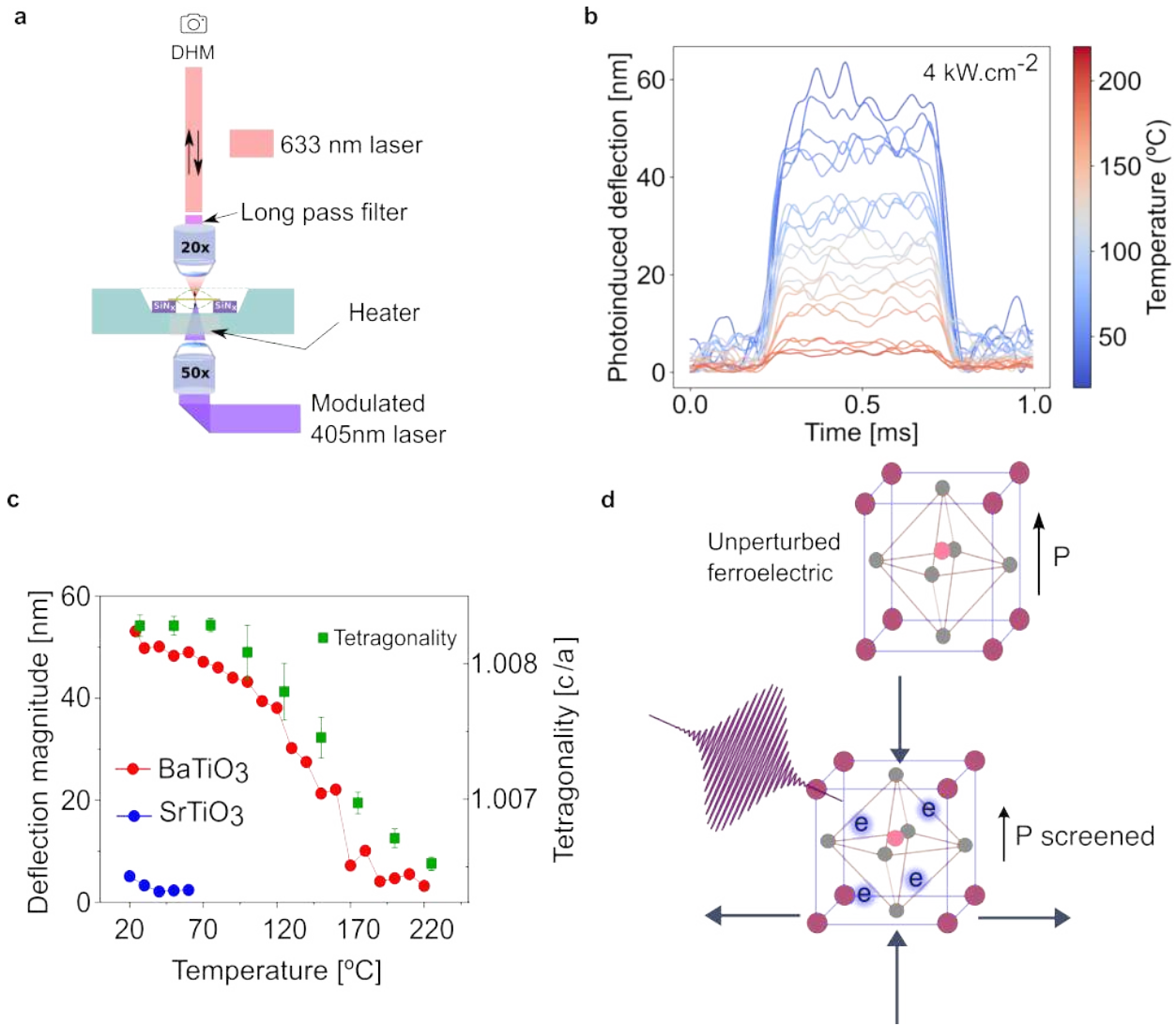

Figure 5: a. The digital holographic microscope setup where the sample is placed in a heater with top and bottom optical access for the 633 nm and the 405 nm laser respectively b. Temperature dependent deflection profiles of a 40nm thin and 20 micron diameter $BaTiO_3$ membrane. c. Deflection magnitude as a function of temperature of $BaTiO_3$ (red) and $SrTiO_3$ (20nm thin and 20 micron diameter, blue) membranes. Photo-deflection of $BaTiO_3$ decreases non-linearly upon heating, mimicking the evolution of its tetragonality (green squares) and therefore spontaneous polarization; implying a direct correlation between photoactuation and ferroelectric polarization. d. Light excitation screens the long range interaction of electric dipoles, weakening the ferroelectric order. Concomitantly, the polar axis contracts and the in-plane (non-polar) axis elongates.

# Supporting Information

# Photostrictive actuators based on freestanding ferroelectric membranes

*Saptam Ganguly[1], David Pesquera[1], Daniel Moreno Garcia[3], Umair Saeed[1],*
*Nona Mirzamohammadi[1], José Santiso[1], Jessica Padilla[1], José Manuel Caicedo Roque[1],*
*Claire Laulhé[4,5], Felisa Berenguer[4], Luis Guillermo Villanueva[3], Gustau Catalan[1,2]*

1. Catalan Institute of Nanoscience and Nanotechnology (ICN2), CSIC and BIST, Campus UAB, Bellaterra, Barcelona, 08193 Spain.

2. ICREA - Institució Catalana de Recerca i Estudis Avançats, Barcelona, Catalonia, 08010 Spain.

3. Advanced NEMS Laboratory, Institute of Mechanical Engineering, École Polytechnique Fédérale de Lausanne (EPFL), 1015, Lausanne, Switzerland.

4. Université Paris-Saclay, Synchrotron SOLEIL, 91190 Saint-Aubin, France.

5. SOLEIL Synchrotron, L'Orme des Merisiers, Saint-Aubin BP 48, 91190 Gif-sur-Yvette, France.

# Section 1:

The Silicon Nitride (Si$_3$N$_4$) substrates are fabricated from double-side polished 100 mm Silicon wafers with 500 nm low-stress Si$_3$N$_4$ (LPCVD deposited). A photoresist layer (3 μm, ECI 3027) is patterned using direct laser writing with the hole designs (a). Si$_3$N$_4$ and Si are etched using Deep Reactive Ion Etching (DRIE) (b). Afterwards, lithography is performed on the backside of the wafer (8 μm, AZ-10XT), aligning with the top side (c). The backside is etched with DRIE following a Bosch® process alternating etching and passivation cycles. The Silicon etching is stopped after 330 μm, therefore leaving 50 μm (d). The photoresist is stripped with 1165 remover and oxygen plasma. The wafer is diced into chips of 5x5 mm. Finally, released BaTiO$_3$ and SrTiO$_3$ epitaxial thin films are transferred onto the chips by using the "pick-up and stamp" method heating the substrate and the attached PDMS stack with the membrane upto 70ºC. This is done to ensure proper adhesion between the thin film and the substrate. (e). For some of the chips, the Silicon under the Si3N4 membrane is completely removed by using 40% KOH for 3h at 60 ºC and ion neutralization with HCl for 2h. Removing the Silicon gives more accessibility for the time-resolved XRD in transmission geometry(f).

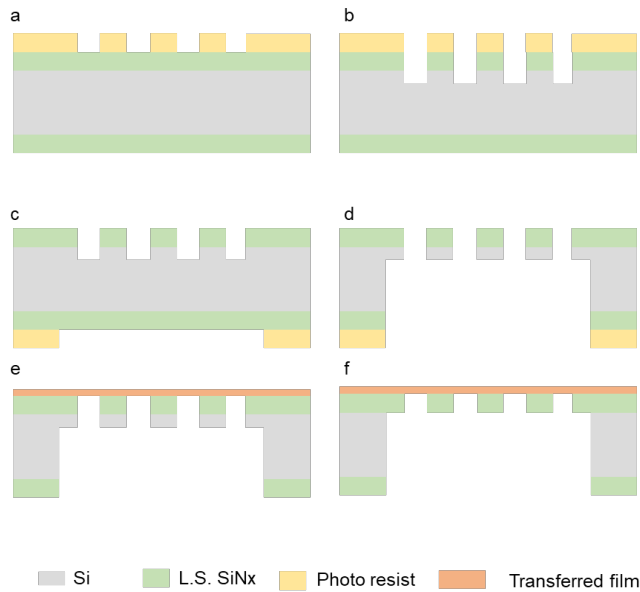

**Fabrication of the substrate.** a, The fabrication starts with silicon wafers with 500 nm Si$_3$N$_4$. The wafer is coated with photoresist and the design is patterned. b, The silicon oxide and the Si are etched with Deep Reactive Ion Etching. c, Lithography to pattern the cavities from the backside. d, DRIE of the Si$_3$N$_4$ and Si, leaving 50 μm of Si. e, Released thin films are transferred on top of the Si$_3$N$_4$ membrane which are then used for Vibrometry measurements. f, For some membranes, Si under the Si$_3$N$_4$ membrane is completely removed using KOH and then thin films are transferred for XRD in transmission.

☐ Si   ☐ L.S. SiNx   ☐ Photo resist   ☐ Transferred film

# Section 2:

The doppler vibrometer uses the doppler effect to study changes of velocity in the object under study. The Doppler effect refers to the change in frequency of light reflected from a moving object, which is proportional to the velocity of the object. Using interferometry to measure this frequency shift, it is possible to precisely determine the vibrational motion of the object. In measurements, involving vibration, the measured frequency shift $\Delta f$ of a laser with wavelength $\lambda_c$ is accurately proportional to the velocity v of the object:

$$\Delta f = 2v/\lambda_c$$

The detected frequency shift is used to determine the velocity v(t), displacement d(t), and acceleration a(t) of the surface from which the laser light is reflected. For harmonic vibrations with frequency 'f' and displacement d(t) = X sin($2\pi ft$), the amplitudes of displacement, velocity, and acceleration are related as follows:

$$A = 2\pi fV = 4\pi^2 f^2 X$$

The frequency changes are converted into an intensity time-series allowing further electronic processing in the frequency domain. Within the interferometer, the laser beam is split into a reference beam and a measurement beam. The light reflected from the object being measured interferes with the reference beam. The intensity recorded by a photodetector includes contributions from the reference beam intensity $I_0$, the reflected light intensity $I_v$, and a term dependent on the difference in optical path $\Delta z$:

$$I(t) = I_0 + I_v + 2(I_0 I_v)^{0.5} \cos\left(\frac{2\pi\Delta z(t)}{\lambda}\right)$$

A heterodyne detection method is employed to distinguish the sign of the movement of the object. By shifting the frequency of the reference beam by a fixed amount $f_d$, the interference of both beams for a stationary object result in harmonic intensity variation at frequency $f_d$.

This carrier signal, proportional to $\cos(2\pi f_d)$, is modulated by the motion of the object. Depending on the direction of motion, the frequency of the intensity is shifted towards higher or lower frequencies. Finally, the intensities are demodulated to get meaningful information. After converting the intensities into a digital signal, a signal processor determines the displacement, velocity, and acceleration of the object in real-time. This processed signal is finally read by a lock-in and then finally converted to real displacement units by a calibration factor provided with the vibrometer.

## Section 3:

Resonant frequency of a plate-like resonator: [1]

$$f_{plate} = \frac{10.21}{4\pi} \sqrt{\frac{E}{3\rho(1-\nu^2)}} \frac{t}{R^2} \quad \text{............................ (1)}$$

where, E is the Young's Modulus, 'ρ' is the density, '$\nu$' is the Poisson Ratio, R is radius of the membrane and t is the thickness.

If the membranes behave as plates, an inverse proportionality to $R^2$ is expected.

Resonant frequency of a membrane-like resonator:

$$f_{membrane} = \frac{2.4048}{2\pi R} \sqrt{\frac{T}{\rho t}} \quad \text{............................ (2)}$$

where, T is the pre-tension in the membrane, 'ρ' is the density, R is radius of the membrane and t is the thickness.

While if a membrane is dictated by the in-plane tension, an inverse proportionality to R is expected.

In cases where both bending rigidity and pre-tension govern, the resonant frequency can be approximated to be,

$$f_0 \approx \sqrt{f_{membrane}^2 + f_{plate}^2} \quad \text{............................ (3)}$$

## Section 4:

Estimate for photocarrier density generated in 40nm thin $BaTiO_3$.

Peak power density used in X-ray microdiffraction experiment = 3 kW.cm$^{-2}$.

Absorbance at 405nm = approx. 0.2

Energy of each photon at 405nm = 3.06 eV * 1.6 * 10$^{-19}$ Joule = 4.89 x 10$^{-19}$ J

Number of photons illuminated per unit area per second (N) =
$\qquad$ Peak power density / energy of each photon
$\qquad$ = 3000 W.cm$^{-2}$ / 4.89 x 10$^{-19}$
$\qquad$ = 6.12 * 10$^{21}$ $\quad$ s$^{-1}$ cm$^{-2}$

Absorbed photon density (A) = absorbance x N
$\qquad\qquad\qquad$ = 1.22 x 10$^{21}$ s$^{-1}$ cm$^{-2}$

Quantum efficiency (QE) of $BaTiO_3$ has been reported to lie in a wide range of values from 2.1 x 10$^{-4}$ to 1 x 10$^{-2}$. [2,3]

Density of e-h pairs created per unit area per second (D) = QE x A,
ranging from 2.57 x 10$^{17}$ to 1.22 x 10$^{19}$ s$^{-1}$ cm$^{-2}$

Density of e-h pairs created per unit volume per second for a 40nm thin film = D / 40 x 10$^{-7}$ (in cm),
ranging from 6.43 x 10$^{22}$ to 3.06 x 10$^{24}$ s$^{-1}$ cm$^{-3}$

Therefore, for the duration of a 500 μs pulse, the number of photocarriers generated = 500 x 10$^{-6}$ s x density of carriers per unit volume, can range from 3.21 x 10$^{19}$ cm$^{-3}$ to 1.53 x 10$^{21}$ cm$^{-3}$.

This lies in the range of predicted carrier density values necessary for 0.047% strain in $BaTiO_3$.

**Figure S1:**

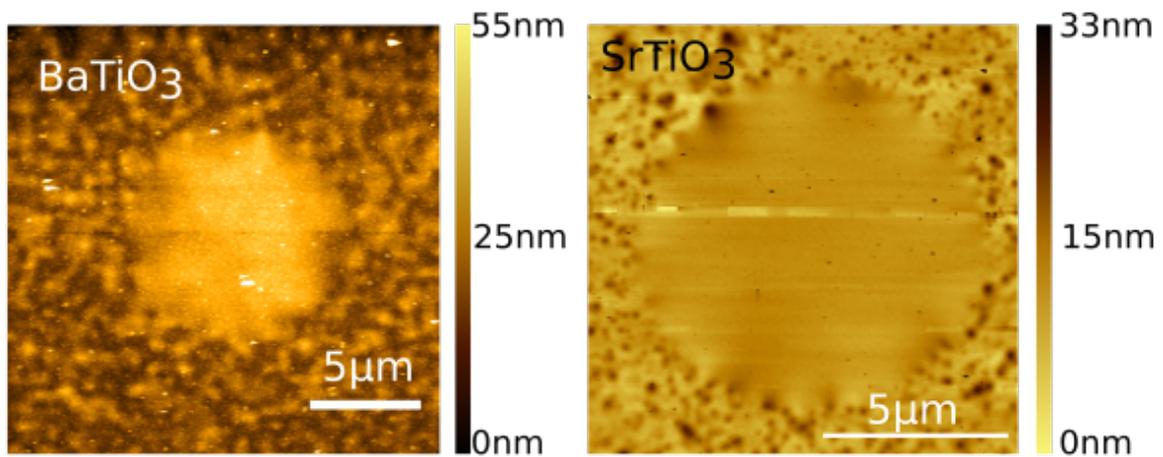

AFM topography of a 40 nm thin BaTiO₃ and 20 nm thin SrTiO₃ membranes of 10 μm diameter. Both BaTiO₃ and SrTiO₃ membranes have a slightly bent state and the surface roughness for both membranes range from 0.2nm to 1.5nm. The surface roughness outside the suspended area is dictated by the roughness of the Silicon Nitride surface present below.

**Figure S2:**

*a.*

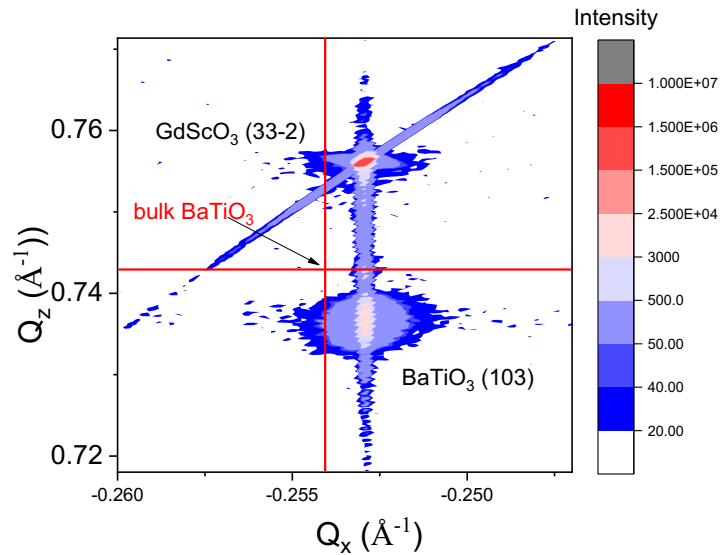

Reciprocal space maps about an asymmetric substrate reflection (33-2) demonstrate that BaTiO$_3$ layer grows coherently strained on the GdScO$_3$ substrate.

*b.*

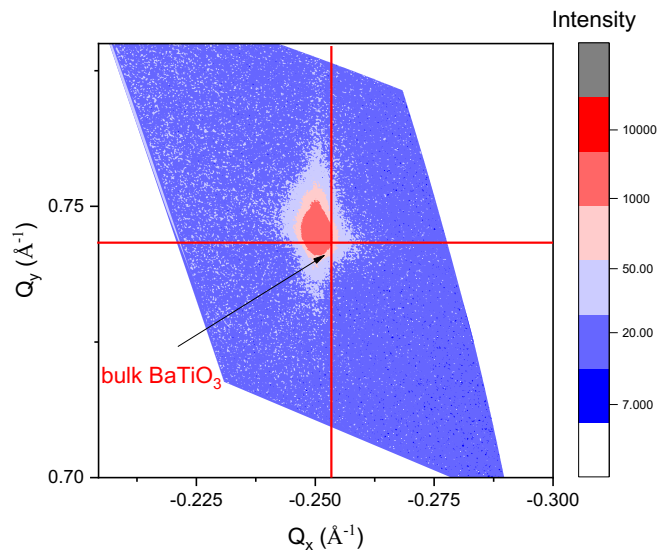

After the sacrificial layer has been dissolved and the membrane has been transferred to a substrate of choice, the reciprocal space maps indicate a relaxation towards bulk lattice parameters of BaTiO$_3$, still withholding the tetragonality.

**Figure S3:**

For comparing mechanical properties of $BaTiO_3$ and $SrTiO_3$ membranes, an indentation experiment was performed whereby both $BaTiO_3$ and $SrTiO_3$ suspended drums of 10μm diameter were pressed in the center with an AFM tip with a fixed loading force range. We observe that for a fixed value of load (for example 200nN) both $BaTiO_3$ and $SrTiO_3$ membranes deform by very similar numbers with a difference of 23nm. The slope (2.68 for $BaTiO_3$ and 2.12 for $SrTiO_3$) of the curve beyond the inflection point, which is proportional to Young's Modulus of the material, indicates very similar mechanical properties. However, in the case of $SrTiO_3$ membrane a nonlinearity is observed which originates from the cubic part of the force-deflection relation.[4]

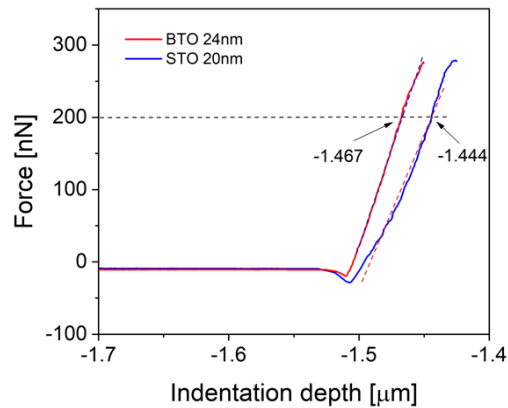

**Figure S4:**

Absorbance (left) comparison of BaTiO₃ (BTO) and SrTiO₃ (STO) both 20nm thick and Tauc plot(right) to determine the direct bandgap. Both BaTiO₃ and SrTiO₃ membranes were transferred on thin block of PDMS. A blank PDMS block was used to acquire a background spectrum and subsequently absorbance spectra for both samples were acquired. To extract the direct bandgap of the materials, the Tauc method was used with an exponent "2" and plotted versus energy. The intercept of the linear slope of the plot onto the energy axis was estimated to be the direct bandgap of the materials. The excitation energy used for the membranes in the experiment (405nm, 3.06eV) lies below the bandgap of both materials.

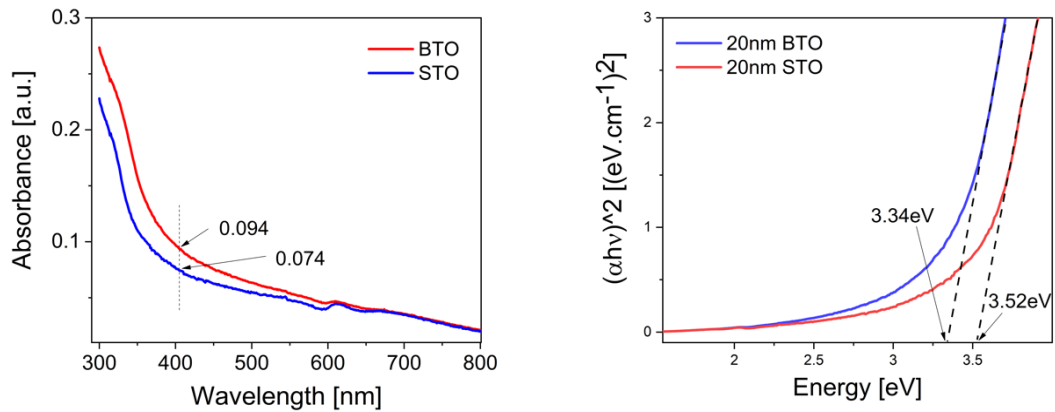

**Figure S5:**

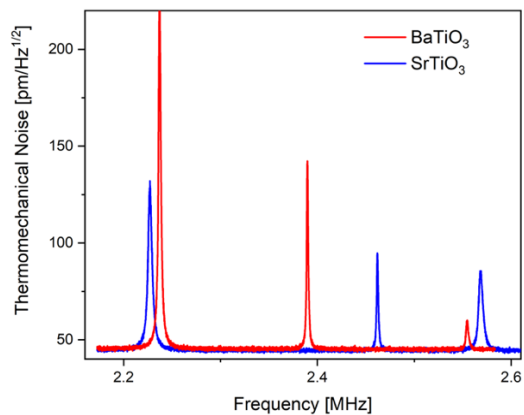

Thermomechanical noise (left) of 30-micron membrane of BaTiO₃ and SrTiO₃, showing multiple resonance modes.

# Figure S6:

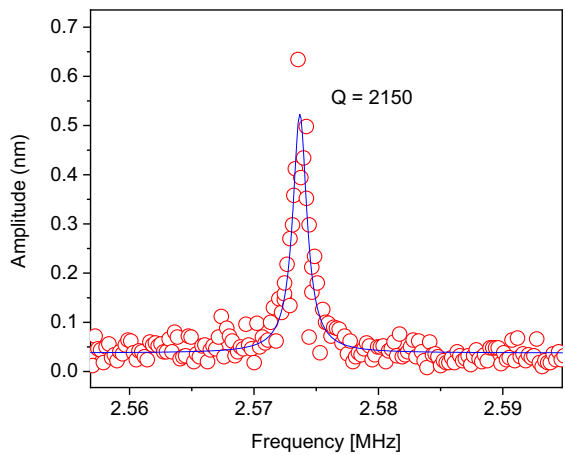

Example of a high Q-factor BaTiO$_3$ membrane 40nm thin and 30 microns in diameter.

# Figure S7:

a. A selected region of interest of the XPAD detector image showing the (200) diffraction spot, from which 2theta- omega profiles are extracted.

delta($\delta$) and gamma($\gamma$) in the detector are related 2$\theta$ as follows:

cos(2$\theta$) = cos($\delta$)*cos($\gamma$). As a function of time the diffraction spot moves along 'delta' axis with a fixed value of gamma and therefore integration along the gamma axis can be used to extract 2theta profiles. The variations in delta and gamma in pixels are converted to degrees by a calibration factor (16millidegree/pixel).

b. Theta- 2theta profiles as a function of delay extracted from the detector image as shown in the previous figure, by integrating on the gamma axis. The peaks were fitted to a pseudo-Voigt function, and the fitted 2$\theta$ values were converted to the change in lattice parameter using Bragg's law.

a.                                    b.

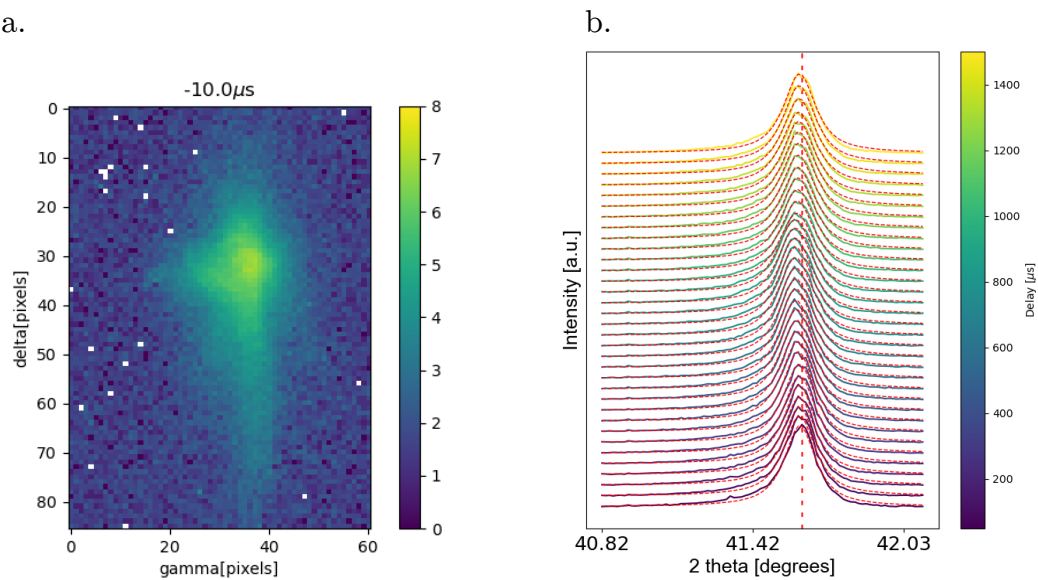

**Figure S8:**

From temperature dependent XRD measurements, the out of plane parameter is extracted by fitting the 2θ-ω curve with a pseudo-voigt function to the peak.

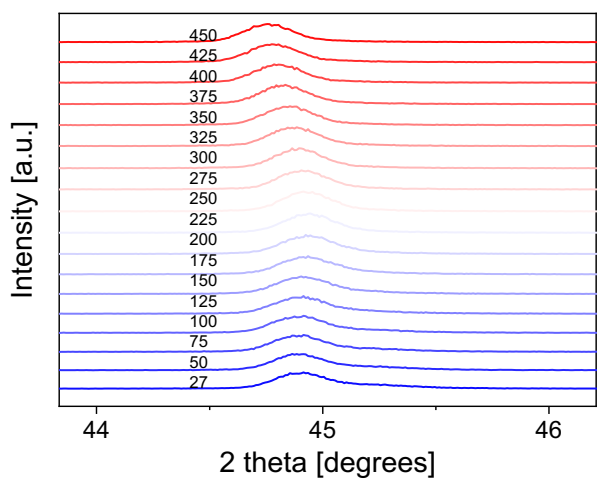

The out of plane lattice parameters in the temperature dependent were extracted by fitting the high-resolution x-ray scans as shown above by fitting a pseudo-Voigt peak while the in-plane lattice parameters were extracted by from the reciprocal space maps (shown below only for two temperatures; 27ºC and 450ºC).

27ºC                                     450ºC

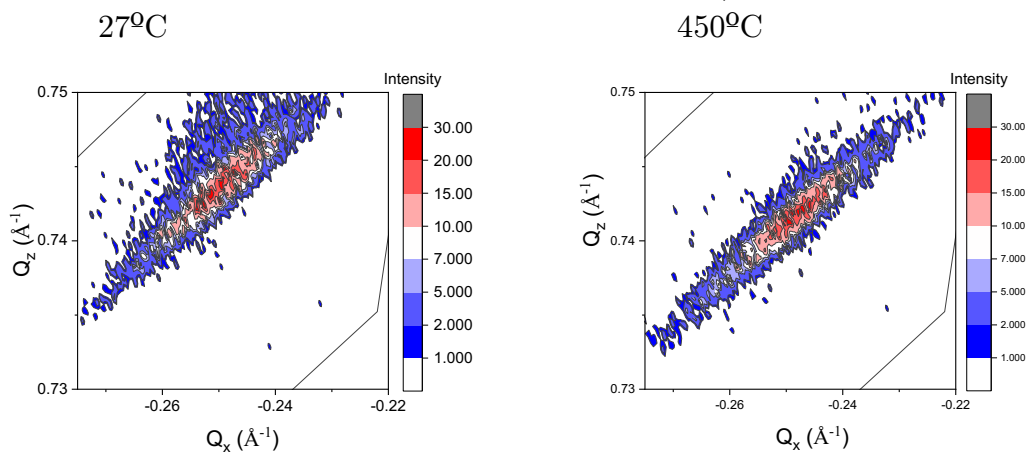

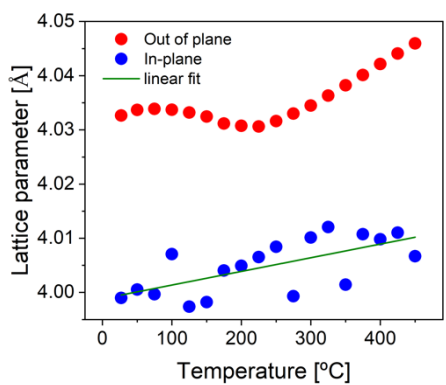

Evolution of out-of-plane and in-plane lattice parameters as a function of temperature. The linear fit to the in-plane lattice parameter data is used to extract the thermal expansion coefficient value. The large deviations in the in-plane lattice parameter data comes from the large spread of the diffraction peak along $Q_x$.

# References:

1.  A. Castellanos-Gomez, V. Singh, S. H. van der Zant, and A. G. Steele. (2015). Mechanics of freely-suspended ultrathin layered materials. Annalen der Physik, 527(1-2), 27-44.
2.  G. Unal, Measurements of the microscopic properties contributing to photorefraction in barium titanate, Technical report, Naval Postgraduate School, Monterey, 1986.
3.  K. H. Yoon, K. S. Chung, Journal of Applied Physics 1992, 72, 12, 5743–5749.
4.  V. Harbola, S. Crossley, S. S. Hong, D. Lu, Y. A. Birkholzer, Y. Hikita and H.Y. Hwang, 2021 Strain gradient elasticity in $SrTiO_3$ membranes: bending versus stretching. Nano Letters, 21(6), 2470-2475.